\documentclass[]{aastex631}
\usepackage{longtable}
\usepackage{footnote}
\usepackage{booktabs}
\usepackage{amsmath}
\usepackage[figuresright]{rotating}
\usepackage{threeparttable}

\shorttitle{Two low mass ratio detached binary}
\shortauthors{Chang et al.}
\graphicspath{{./}{figures/}}

\begin{document}
\correspondingauthor{Shengbang Qian}
\email{qiansb@ynu.edu.cn}

\title{Detection of two totally eclipsing B-type binaries with extremely low mass ratios}

\author[0000-0002-8421-4561]{Linfeng Chang}
\affiliation{Department of Astronomy, School of Physics and Astronomy, Yunnan University, Kunming 650091, China}
\affiliation{Key Laboratory of Astroparticle Physics of Yunnan Province, Yunnan University, Kunming 650091, China}

\author{Shengbang Qian}
\affiliation{Department of Astronomy, School of Physics and Astronomy, Yunnan University, Kunming 650091, China}
\affiliation{Key Laboratory of Astroparticle Physics of Yunnan Province, Yunnan University, Kunming 650091, China}

\author{Lei Zang}
\affiliation{Department of Astronomy, School of Physics and Astronomy, Yunnan University, Kunming 650091, China}
\affiliation{Key Laboratory of Astroparticle Physics of Yunnan Province, Yunnan University, Kunming 650091, China}

\author{Fuxing Li}
\affiliation{Department of Astronomy, School of Physics and Astronomy, Yunnan University, Kunming 650091, China}
\affiliation{Key Laboratory of Astroparticle Physics of Yunnan Province, Yunnan University, Kunming 650091, China}

\begin{abstract}
The detection of O- and B-type stars with extremely low-mass companions is very important for understanding the formation and evolution of binary stars. However, their finding remains a challenge because the low-mass components in such systems contribute such small flux to the total. During the searching for pulsations among O- and B-type stars by using the TESS data, we found two short-period and B-type (B9) eclipsing binaries with orbital periods of 1.61613 and 2.37857 days. Photometric solutions of the two close binaries were derived by analyzing the TESS light curves with the W-D method. It is discovered that both of them are detached binaries with extremely low mass ratios of 0.067(2) for TIC 260342097 and 0.140(3) for TIC 209148631, respectively. The determined mass ratio indicates that TIC 260342097 is one of the lowest mass ratios among known B-type binary systems. We showed that the two systems have total eclipses with a broad and flat secondary minimum, suggesting that the photometric parameters could be derived reliably. The absolute parameters of the two binaries are estimated and it is found that the secondary components in the two systems are over-luminous and over-size when compared with the normal low-mass and cool main-sequence (MS) stars. These findings may imply that the two systems are composed of a B-type MS primary and a cool pre-MS secondary with orbital periods shorter than 2.5 days. They are valuable targets to test theories of binary star formation and evolution.

\end{abstract}


\section{Introduction} \label{sec:intro}

 O- and B- type stars often exist in binary or multiple systems, and the fraction of binary or multiple systems increases with the stellar mass. \citet{2014ApJS..215...15S} observed hundreds of O-type stars, finding the proportion of samples that exhibit at least one resolved companion within a separation of 200 mas is 0.53. After considering known but unresolved spectroscopic or eclipsing companions, \citet{2014ApJS..215...15S} pointed out that the multiplicity fraction at separation under $8^{\prime\prime}$ increases to 91\%. The statistics from \citet{2017ApJS..230...15M} also support the notion of a high frequency of O- or B-type stars having at least one companion. However, the multiplicity fraction of massive stars maybe be underestimated owning to the difficulties in detection of their low-mass companions, because the luminosity contribution from the secondary component is too small to be detected \citep{2011IAUS..272..474S, 2013AJ....145....3G}. Therefore, \citet{2014ApJS..215...15S} deduce that the binary ratio of massive stars of V class is at virtually 100\%, thus robustly demonstrating that massive stars predominantly form in multiple systems. Searching for low-mass companions to massive stars is very important to understand the binary fraction of O- and B- type stars. 
 
 According to the statistic of \citet{2019MmSAI..90..359B}, B-type stars display an excess of low-mass companions. Binaries with low mass ratio are significant probes of stellar formation due to the way they form may be different from high mass-ratio ones. Currently, the primary way of binary formation is thought to be core fragmentation, where a collapsing protostellar core that produces two or more very dense parts which then separately begin collapsing \citep{1995MNRAS.277..362B}. In addition, fragmentation of a massive protostar’s disc is another important way to form binary systems. In this scenario, the accretion disc of a massive star is likely to fragment due to gravitational instability \citep{2006MNRAS.373.1563K}. \citet{2011ApJ...730...32S} pointed out that episodic accretion of a protostar drives the formation of low-mass stars by disk fragmentation. While \citet{2013AJ....145....3G} suggested that disk fragmentation typically yields lower-mass companions than core fragmentation. Investigations of the low mass-ratio binary systems can shed light on the comparative importance of both mechanisms in forming high-mass binary systems, and may contribute to refining models as computational capabilities advance.

\begin{table}
    \centering
     \caption{The basic information of TIC 260342097 and TIC 209148631.}\label{basic-inf}
    \begin{tabular}{lccccc}
\hline
       Name (TIC)    &  Right Ascension ($\alpha_{2000}$) &  Declination ($\delta_{2000}$)& Mag (V) & Parallax (mas) & Spectral type  \\
 \hline
      260342097 &15:00:08.8 &-65:59:07.8  &  9.86 &2.3895 &B9,B9.5V \\
      209148631 &13:59:23.3 & -59:02:16.1 &  10.51& 1.1441 & B8/9V,A0\\
\hline
\end{tabular}  
\end{table}

TESS (Transiting Exoplanet Survey Satellite) is originally constructed to search for exoplanets orbiting around nearby bright stars \citep{2015JATIS...1a4003R}. However, the continuously time-series photometric data obtained by TESS are very useful to study pulsating stars and eclipsing binaries. Based on TESS data, we are recently searching for and investigating slowly pulsating B-type stars as well as $\beta$ Cephei variable stars among O- and B-type massive stars \citep{2023ApJS..268...16S,2023ApJS..265...33S}. During the works, two B-type stars (TIC 260342097 and TIC 209148631) were found to be eclipsing binary systems. TIC 260342097 (or HD 131987 and TYC 9032-822-1), a member of the open star cluster BH\_164\citep{2018A&A...618A..93C}, was cataloged as a star in the Simbad database\footnote{\url{http://simbad.u-strasbg.fr/simbad/sim-fid}}. 
Its spectral type was classified as B9.5V by \citet{1975mcts.book.....H} and B9 from \citet{1993yCat.3135....0C}. However, the other star, TIC 209148631 (or HD121793 and TYC8676-3476-1), was listed as a non-variable source in the TESS Input Catalog by \citet{2018AJ....155...39O} based on observations from the Kilodegree Extremely Little Telescope (KELT). It was later identified as a pulsating variable star by \citet{2022yCat.1358....0G}. \citet{1975mcts.book.....H} and \citet{1993yCat.3135....0C} classified its spectral type as B8/9V and A0, respectively. In the work of \citet{2023MNRAS.522...29G}, orbital period estimates are provided for both TIC 260342097 and TIC 209148631 when they conducted a search for ellipsoidal variable candidates. The basic information of the two targets were compiled in Table \ref{basic-inf}, where the parallaxes are from Gaia \citep{2023A&A...674A...1G}. To date, no study has been conducted to determine crucial parameters, such as mass ratios and orbital inclinations, for these two binaries.

\section{Data Acquisition and Reduction}  \label{TESSdata}

\begin{figure}[htbp]
\centering
\includegraphics[width=10cm]{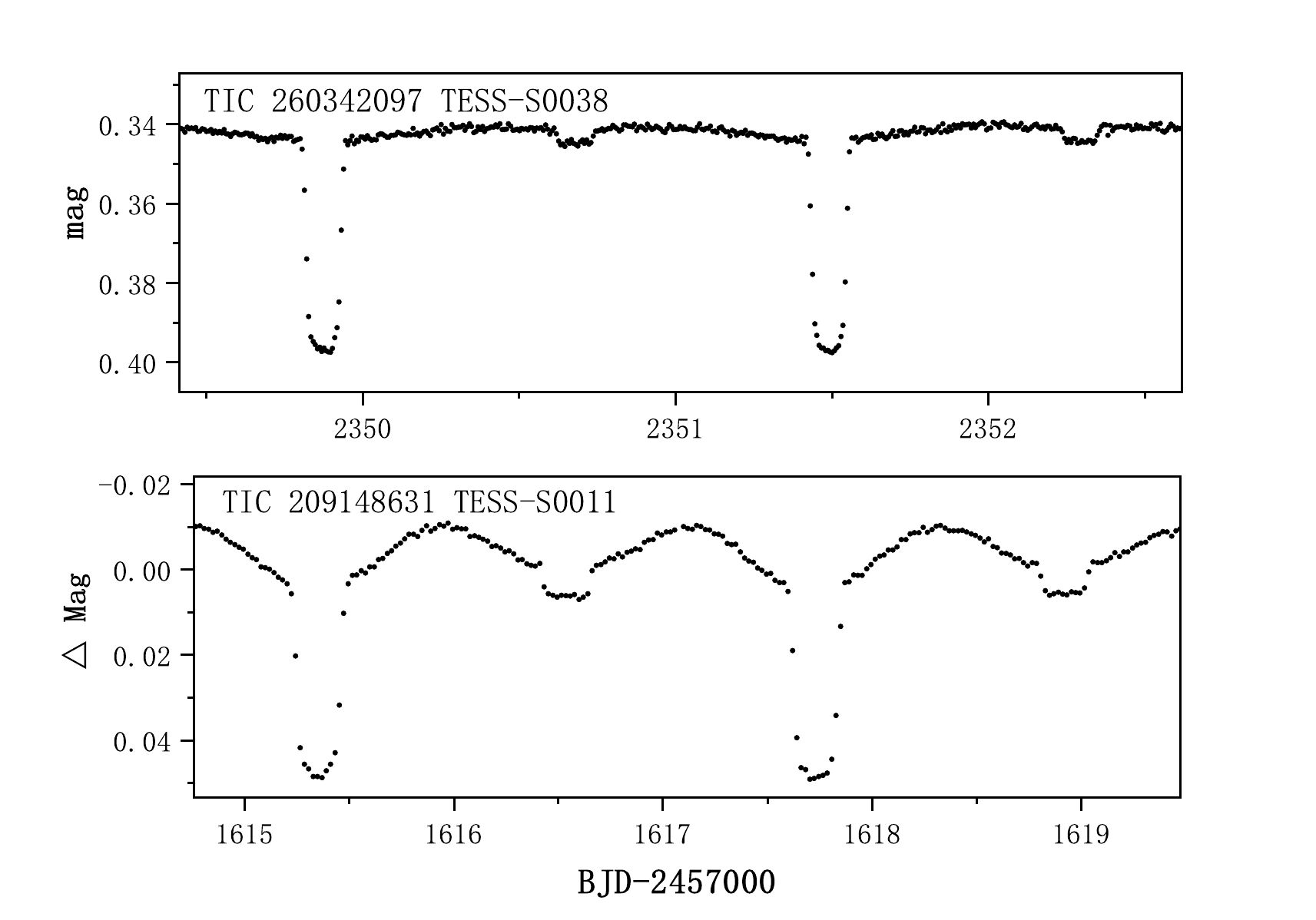}
\caption{The light curves of TIC 260342097 and TIC 209148631 from TESS.}
\label{time-LC}
\end{figure}

TIC 260342097 was observed by the TESS space mission for four times. Its photometric data were archived at the Mikulski Archive for Space Telescopes (MAST). We downloaded all data from this portal, which are sectors 12, 38, 39 and 65 with the exposure times of 1800S, 600S, 600S and 200S, respectively. All light curves used were processed by the Science Processing Operations Center (SPOC, \citealt{2016SPIE.9913E..3EJ}), and the Pre-search Data Conditioning Simple Aperture Photometry (PDC SAP) flux was adopted. While for TIC 209148631, only sector 11 with a cadence of 30 min is available, and we applied the PDC SAP flux processed by the SPOC to carry out the analysis. Here we converted all the flux into magnitude with $mag=-2.5\times log(flux)$ because the relative changes of the light curves are more concerned about. Given that there were no discernible discrepancies among the four-sector data of TIC 260342097, the TESS data of sector 38 were adopted to carry out photometric analysis, meanwhile, the analysis on TIC 209148631 was based on the data of sector 11. Figure \ref{time-LC} illustrates the light curves of the two objects, each demonstrating a notably broad and flat secondary minimum, manifesting that the secondary eclipses are total eclipses. Thus, the fractional sum of radii of TIC 260342097 and TIC 209148631 can be well constrained and we can derive reliable photometric elements of them \citep{2005Ap&SS.296..221T}.

\section{Determination of the orbital period of TIC 260342097 and TIC 209148631} \label{period-re}

\begin{figure}[htbp]
\begin{minipage}[t]{0.49\textwidth}
\centering
\includegraphics[width=8cm]{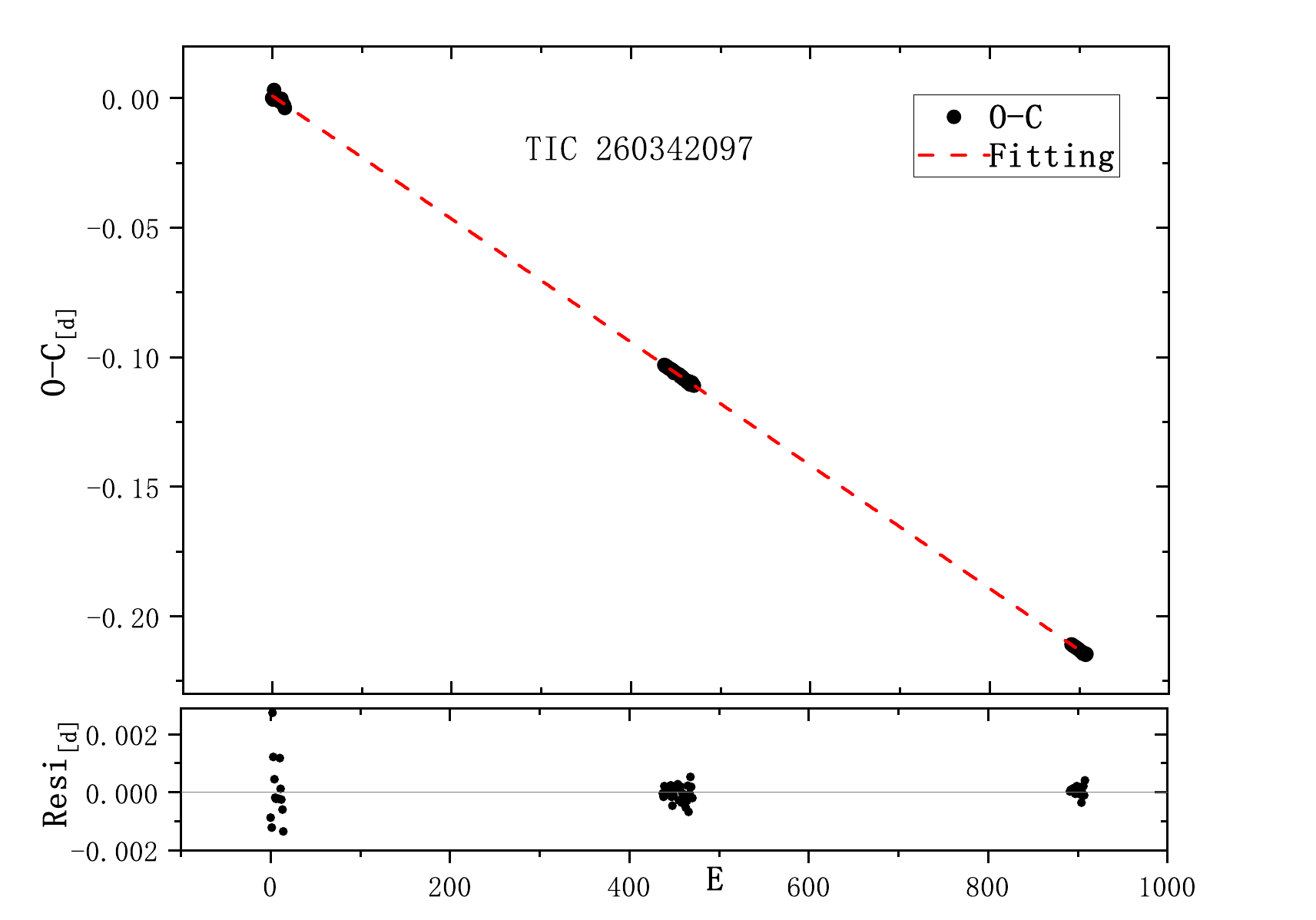}
\end{minipage}
\begin{minipage}[t]{0.49\textwidth}
\centering
\includegraphics[width=8cm]{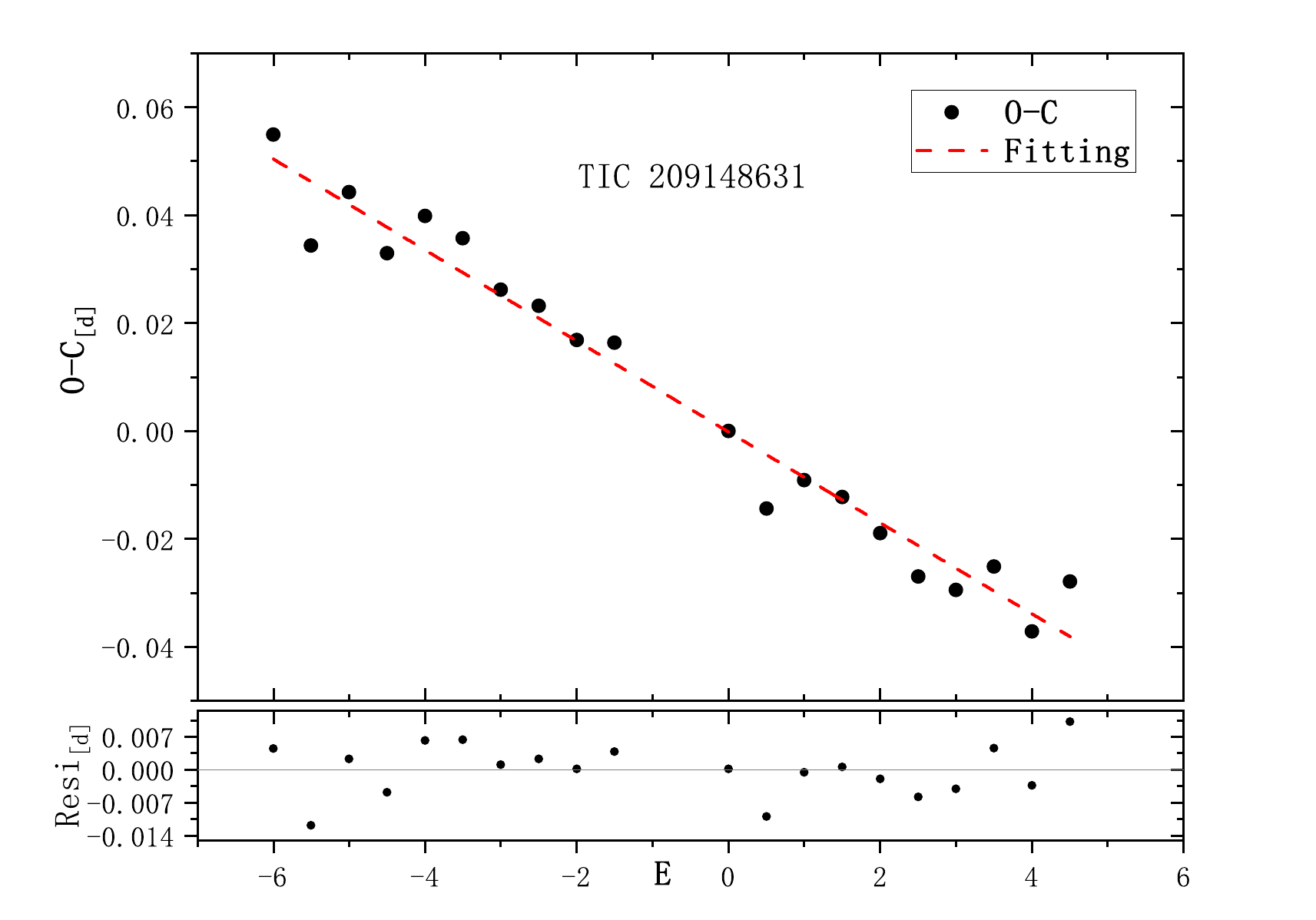}
\end{minipage}
\caption{$O-C$ diagram of TIC 260342097 (left panel) and TIC 209148631 (right panel), which are used to improve the initial periods from Period04, thus finally determining the orbital periods of both binaries.}
\label{fig:OC}
\end{figure}

The orbital period carries vital insights into the characteristics of binary stars. The periods of TIC 260342097 and TIC 209148631 were reported as 0.806334 days and 1.191560 days from \citet{2023MNRAS.522...29G}, respectively. And period of TIC 209148631 was also reported as 1.189187 days from \citet{2022yCat.1358....0G}. However, upon initial inspection of the TESS data, we found that the previously reported periods were incorrect. In order to derive accurate periods for these two systems, the Period04 software \citep{2005CoAst.146...53L} was first applied to their light curves of TESS to derive a initial orbital period, by which P=1.61637(58) days and P=2.3870(29) days were obtained for TIC 260342097 and TIC 209148631. This work is the first detailed study on TIC 260342097 and TIC 209148631, and no existing eclipse timings available in the literature yet. Furthermore, photometric data from other publicly accessible resources, such as the American Association of Variable Star Observers (AAVSO), Super Wide Angle Search for Planets (SuperWASP; \citealt{2006PASP..118.1407P}), Catalina Sky Survey \citep{2014ApJS..213....9D} and All-Sky Automated Survey for Supernovae (ASAS-SN; \citealt{2017PASP..129j4502K}), were also examined. Among these sources, only ASAS-SN provided photometric observations for these two targets; however, the data were too scattered to reliably derive eclipse timings. Consequently, our period analyses rely solely on the photometric data from TESS. The observations of sectors 12, 38, 39 and 65 were used to acquire the eclipse timings of TIC 260342097 using the method of \cite{1956BAN....12..327K}, from which 59 primary minima (P) and 58 secondary minima (S) were determined. All eclipsing timings were tabulated in Table \ref{tab:o-c}. However, considering the secondary minima of TIC 260342097 are too shallow to derive accurate eclipse timings, only the primary minima were used to improve the initial period. The O-C values were computed with the following ephemeris:
\begin{equation}
\label{eq-oc1}
     Min I=BJD 2458629.07859+1^{d}.61637 \times E 
\end{equation}
According to Eqn (\ref{eq-oc1}), the cycle numbers (E) and O-C values, along with these of secondary minima, were computed and listed in Table \ref{tab:o-c}. The corresponding O-C trend was displayed in the left panel of Figure \ref{fig:OC}, which showed an obvious linear distribution, we thus adopted a linear equation to fit the O-C values, revising a new ephemeris of TIC 260342097 as:
\begin{equation}
\label{eq-oc2}
     Min I=BJD 2458629.07946 (13)+1^{d}.6161323(2) \times E 
\end{equation}

Despite the fact that data of merely one sector for TIC 209148631 were available from TESS, we endeavored to obtain a more precise period of it. Observations from sector 11 were utilized, from which 10 primary minima and 10 secondary minima were extracted. These minima, together with E and corresponding O-C values from Eqn (\ref{eq-oc4}), were presented in the last 10 rows of Table \ref{tab:o-c}. Then, the 20 eclipsing timings were applied to construct O-C diagram for TIC 209148631. The O-C values were calculated with the following ephemeris:

\begin{equation}
\label{eq-oc4}
     Min I=BJD 2458612.97541+2^{d}.387 \times E 
\end{equation}
The corresponding O-C diagram, displayed in the right of Figure \ref{fig:OC}, also showed a linear distribution, and a linear equation was employed to fit the O-C values, deriving an updated ephemeris of TIC 209148631 as:
\begin{equation}
\label{eq-oc2}
     Min I=BJD 2458612.97527(130)+2^{d}.37857(40) \times E 
\end{equation}
However, one can notice that the eclipse timings employed to refine the period of TIC 209148631 cover a limited time span. Therefore, space photometric data of TIC 209148631 in the future are necessary to ascertain a more precise orbital period.

\begin{center}
\begin{normalsize}
\begin{longtable}{cccccccccc}
\caption{Times of light minimum of TIC 260342097 and TIC 209148631.}\label{tab:o-c} \\
\hline
BJD       & error   & E   & O-C   &P/S & BJD       & error   & E   & O-C   &P/S\\
2457000+  &  days   &     & days  &     &2457000+  &  days   &     & days   &  \\
\hline \endfirsthead
\multicolumn{10}{l}{Continued table \ref{tab:o-c}}\\ \hline
BJD       & error   & E   & O-C   &P/S & BJD       & error   & E   & O-C   &P/S\\
2457000+  &  days   &     & days  &     &2457000+  &  days   &     & days   &  \\
\hline \endhead
\hline \endfoot
1628.24455 	&	0.00004 	&	-0.5	&	-0.025855 	&	S	&	2365.22578 	&	0.00003 	&	455.5	&	-0.109345 	&	S	\\
1629.07859 	&	0.00118 	&	0	&	0.000000 	&	P	&	2366.03590 	&	0.00062 	&	456	&	-0.107410 	&	P	\\
1629.88773 	&	0.00014 	&	0.5	&	0.000955 	&	S	&	2366.84142 	&	0.00016 	&	456.5	&	-0.110075 	&	S	\\
1630.69437 	&	0.00032 	&	1	&	-0.000590 	&	P	&	2367.65212 	&	0.00046 	&	457	&	-0.107560 	&	P	\\
1631.50958 	&	0.00015 	&	1.5	&	0.006435 	&	S	&	2368.45655 	&	0.00003 	&	457.5	&	-0.111315 	&	S	\\
1632.31447 	&	0.00015 	&	2	&	0.003140 	&	P	&	2369.26769 	&	0.00019 	&	458	&	-0.108360 	&	P	\\
1633.11585 	&	0.00002 	&	2.5	&	-0.003665 	&	S	&	2370.07285 	&	0.00011 	&	458.5	&	-0.111385 	&	S	\\
1633.92907 	&	0.00107 	&	3	&	0.001370 	&	P	&	2370.88415 	&	0.00056 	&	459	&	-0.108270 	&	P	\\
1634.73278 	&	0.00011 	&	3.5	&	-0.003105 	&	S	&	2371.68730 	&	0.00007 	&	459.5	&	-0.113305 	&	S 	\\
1635.54444 	&	0.00022 	&	4	&	0.000370 	&	P	&	2372.50007 	&	0.00031 	&	460	&	-0.108720 	&	P	\\
1636.34937 	&	0.00002 	&	4.5	&	-0.002885 	&	S	&	2373.30387 	&	0.00002 	&	460.5	&	-0.113105 	&	S	\\
1637.15993 	&	0.00018 	&	5	&	-0.000510 	&	P	&	2374.11630 	&	0.00009 	&	461	&	-0.108860 	&	P	\\
1638.00453 	&	0.00047 	&	5.5	&	0.035905 	&	S	&	2376.53503 	&	0.00009 	&	462.5	&	-0.114685 	&	S 	\\
1638.77603 	&	0.00023 	&	6	&	-0.000780 	&	P	&	2377.34818 	&	0.00027 	&	463	&	-0.109720 	&	P	\\
1642.81603 	&	0.00008 	&	8.5	&	-0.001705 	&	S	&	2378.15210 	&	0.00005 	&	463.5	&	-0.113985 	&	S	\\
1643.62443 	&	0.00064 	&	9	&	-0.001490 	&	P	&	2378.96455 	&	0.00059 	&	464	&	-0.109720 	&	P	\\
1644.42885 	&	0.00008 	&	9.5	&	-0.005255 	&	S	&	2379.76975 	&	0.00011 	&	464.5	&	-0.112705 	&	S	\\
1645.24196 	&	0.00028 	&	10	&	-0.000330 	&	P	&	2380.58120 	&	0.00035 	&	465	&	-0.109440 	&	P	\\
1646.03880 	&	0.00005 	&	10.5	&	-0.011675 	&	S	&	2381.38531 	&	0.00006 	&	465.5	&	-0.113515 	&	S	\\
1646.85704 	&	0.00032 	&	11	&	-0.001620 	&	P	&	2382.19643 	&	0.00015 	&	466	&	-0.110580 	&	P	\\
1647.65993 	&	0.00007 	&	11.5	&	-0.006915 	&	S	&	2382.99390 	&	0.00013 	&	466.5	&	-0.121295 	&	S	\\
1648.47280 	&	0.00055 	&	12	&	-0.002230 	&	P	&	2383.81319 	&	0.00070 	&	467	&	-0.110190 	&	P	\\
1649.27772 	&	0.00006 	&	12.5	&	-0.005495 	&	S	&	2384.61972 	&	0.00010 	&	467.5	&	-0.111845 	&	S	\\
1650.08858 	&	0.00033 	&	13	&	-0.002820 	&	P	&	2385.42990 	&	0.00032 	&	468	&	-0.109850 	&	P	\\
1650.89139 	&	0.00014 	&	13.5	&	-0.008195 	&	S	&	2386.23372 	&	0.00004 	&	468.5	&	-0.114215 	&	S 	\\
1651.70396 	&	0.00086 	&	14	&	-0.003810 	&	P	&	2387.04570 	&	0.00080 	&	469	&	-0.110420 	&	P	\\
1652.50766 	&	0.00007 	&	14.5	&	-0.008295 	&	S	&	2387.84755 	&	0.00015 	&	469.5	&	-0.116755 	&	S	\\
2334.52166 	&	0.00006 	&	436.5	&	-0.102435 	&	S	&	2388.66144 	&	0.00047 	&	470	&	-0.111050 	&	P	\\
2335.32923 	&	0.00023 	&	437	&	-0.103050 	&	P	&	2389.46430 	&	0.00007 	&	470.5	&	-0.116375 	&	S	\\
2336.13343 	&	0.00002 	&	437.5	&	-0.107035 	&	S	&	3069.05337 	&	0.00014 	&	891	&	-0.210890 	&	P	\\
2336.94525 	&	0.00059 	&	438	&	-0.103400 	&	P	&	3069.81324 	&	0.00101 	&	891.5	&	-0.259205 	&	S	\\
2337.75461 	&	0.00013 	&	438.5	&	-0.102225 	&	S	&	3070.66955 	&	0.00017 	&	892	&	-0.211080 	&	P	\\
2338.56175 	&	0.00046 	&	439	&	-0.103270 	&	P	&	3072.28570 	&	0.00014 	&	893	&	-0.211300 	&	P	\\
2339.36354 	&	0.00008 	&	439.5	&	-0.109665 	&	S	&	3073.08762 	&	0.00097 	&	893.5	&	-0.217565 	&	S	\\
2340.17770 	&	0.00010 	&	440	&	-0.103690 	&	P 	&	3073.90177 	&	0.00011 	&	894	&	-0.211600 	&	P	\\
2340.98135 	&	0.00010 	&	440.5	&	-0.108225 	&	S 	&	3074.67207 	&	0.00148 	&	894.5	&	-0.249485 	&	S	\\
2341.79381 	&	0.00061 	&	441	&	-0.103950 	&	P	&	3075.51800 	&	0.00015 	&	895	&	-0.211740 	&	P	\\
2342.59852 	&	0.00008 	&	441.5	&	-0.107425 	&	S	&	3076.34988 	&	0.00096 	&	895.5	&	-0.188045 	&	S	\\
2343.40985 	&	0.00026 	&	442	&	-0.104280 	&	P	&	3077.13414 	&	0.00017 	&	896	&	-0.211970 	&	P	\\
2344.21759 	&	0.00009 	&	442.5	&	-0.104725 	&	S	&	3077.96782 	&	0.00419 	&	896.5	&	-0.186475 	&	S	\\
2345.02616 	&	0.00081 	&	443	&	-0.104340 	&	P	&	3078.75008 	&	0.00016 	&	897	&	-0.212400 	&	P	\\
2345.83149 	&	0.00008 	&	443.5	&	-0.107195 	&	S	&	3079.57428 	&	0.00089 	&	897.5	&	-0.196385 	&	S	\\
2347.44715 	&	0.00004 	&	444.5	&	-0.107905 	&	S	&	3080.36633 	&	0.00089 	&	898	&	-0.212520 	&	P	\\
2348.25837 	&	0.00024 	&	445	&	-0.104870 	&	P	&	3081.15610 	&	0.00105 	&	898.5	&	-0.230935 	&	S	\\
2349.06120 	&	0.00024 	&	445.5	&	-0.110225 	&	S	&	3081.98261 	&	0.00015 	&	899	&	-0.212610 	&	P	\\
2349.87470 	&	0.00067 	&	446	&	-0.104910 	&	P	&	3084.41448 	&	0.00053 	&	900.5	&	-0.205295 	&	S	\\
2350.67865 	&	0.00018 	&	446.5	&	-0.109145 	&	S	&	3085.21462 	&	0.00016 	&	901	&	-0.213340 	&	P	\\
2351.49044 	&	0.00037 	&	447	&	-0.105540 	&	P	&	3086.01993 	&	0.00041 	&	901.5	&	-0.216215 	&	S	\\
2352.29630 	&	0.00004 	&	447.5	&	-0.107865 	&	S	&	3086.83085 	&	0.00013 	&	902	&	-0.213480 	&	P	\\
2353.10627 	&	0.00079 	&	448	&	-0.106080 	&	P	&	3087.65167 	&	0.00078 	&	902.5	&	-0.200845 	&	S	\\
2353.90624 	&	0.00010 	&	448.5	&	-0.114295 	&	S	&	3088.44685 	&	0.00014 	&	903	&	-0.213850 	&	P	\\
2354.72276 	&	0.00048 	&	449	&	-0.105960 	&	P	&	3090.06270 	&	0.00017 	&	904	&	-0.214370 	&	P	\\
2355.52734 	&	0.00006 	&	449.5	&	-0.109565 	&	S	&	3091.67918 	&	0.00015 	&	905	&	-0.214260 	&	P	\\
2356.33910 	&	0.00027 	&	450	&	-0.105990 	&	P	&	3092.46850 	&	0.00074 	&	905.5	&	-0.233125 	&	S	\\
2357.14020 	&	0.00002 	&	450.5	&	-0.113075 	&	S	&	3093.29553 	&	0.00014 	&	906	&	-0.214280 	&	P	\\
2361.99030 	&	0.00010 	&	453.5	&	-0.112085 	&	S	&	3094.12315 	&	0.00092 	&	906.5	&	-0.194845 	&	S	\\
2362.80380 	&	0.00064 	&	454	&	-0.106770 	&	P	&	3094.91135 	&	0.00017 	&	907	&	-0.214830 	&	P	\\
2363.61085 	&	0.00008 	&	454.5	&	-0.107905 	&	S	&	3096.52800 	&	0.00021 	&	908	&	-0.214550 	&	P	\\
2364.41938 	&	0.00024 	&	455	&	-0.107560 	&	P	\\											1598.70830 	&	0.00348 	&	-6	&	0.05489 	&	P	&	1612.97541 	&	0.00418 	&	0	&	0.00000 	&	P	\\
1599.88127 	&	0.00996 	&	-5.5	&	0.03436 	&	S	&	1614.15458 	&	0.00675 	&	0.5	&	-0.01433 	&	S	\\
1601.08466 	&	0.00296 	&	-5	&	0.04425 	&	P	&	1615.35330 	&	0.00330 	&	1	&	-0.00911 	&	P	\\
1602.26687 	&	0.00440 	&	-4.5	&	0.03296 	&	S	&	1616.54371 	&	0.00817 	&	1.5	&	-0.01220 	&	S	\\
1603.46722 	&	0.00347 	&	-4	&	0.03981 	&	P	&	1617.73049 	&	0.00305 	&	2	&	-0.01892 	&	P	\\
1604.65665 	&	0.00470 	&	-3.5	&	0.03574 	&	S	&	1618.91595 	&	0.00482 	&	2.5	&	-0.02696 	&	S	\\
1605.84060 	&	0.00336 	&	-3	&	0.02619 	&	P	&	1620.10695 	&	0.00304 	&	3	&	-0.02946 	&	P	\\
1607.03111 	&	0.00581 	&	-2.5	&	0.02320 	&	S	&	1621.30483 	&	0.00400 	&	3.5	&	-0.02508 	&	S	\\
1608.21827 	&	0.00338 	&	-2	&	0.01686 	&	P	&	1622.48626 	&	0.00324 	&	4	&	-0.03715 	&	P	\\
1609.41125 	&	0.00544 	&	-1.5	&	0.01634 	&	S	&	1623.68903 	&	0.00587 	&	4.5	&	-0.02788 	&	S	\\

\hline
\end{longtable}
\end{normalsize}
\begin{tablenotes}
\small
The last 10 rows are eclipse timings of TIC 209148631. 
\end{tablenotes}
\end{center}

\section{Light curve analysis} \label{W-D}
The light curves play a critical role in revealing valuable information of eclipsing binaries. TIC 209148631 was recognized as a pulsating variable star by \citet{2022yCat.1358....0G}, who reported a $\sim$ 1.19-day period seeing in the Gaia DR3 G-band data. In order to check for the existence of pulsation signals in TIC 260342097 and TIC 209148631, we employed the Period04 software to investigate the presence of any additional periodic signals beyond the binary orbits in both targets. Their TESS data were segmented into various intervals (e.g., single cycles, 1-day segments, individual sectors, or combined across all sectors) to conduct analysis. In each attempt, no reliable signals other than those associated with the eclipse period and its harmonic frequencies were detected. To specifically verify the 1.19-day period noted by \citet{2022yCat.1358....0G}, we conducted a frequency analysis using the G-band photometric data from Gaia DR3 for TIC 209148631. The result confirmed that the frequency at approximately 1.19 days is a harmonic frequency of the eclipse frequency ($\sim$ 1/2.38 $d^{-1}$) from TESS data. Consequently, we conclude that the variability reported by the \citet{2022yCat.1358....0G} for TIC 209148631 is indeed attributable to the eclipse phenomenon. Thus, our analyses indicate an absence of pulsation signals with the light curves of both TIC 260342097 and TIC 209148631.

To derive the photometric results of the two binaries, the Wilson-Devinney (W–D) code \citep{1971ApJ...166..605W,1979ApJ...234.1054W} was applied to analyze the light curve of TIC 260342097 and TIC 209148631 from TESS. In the analysis process, the light curves covering two full orbits for TIC 260342097 and three for TIC 209148631 were folded into phase. The phases were calculated with the improved periods from Section \ref{period-re}:
\begin{equation} 
    Min.I(BJD)=2459351.49085+1.^{d}6161323\times E
\label{equ1}
\end{equation}
and 
\begin{equation} 
Min.I(BJD)=2458617.72954+2.^{d}37857\times E
\label{equ2}
\end{equation}
Then we binned the folded light curve of TIC 209148631 with an interval of 0.003 in phase, while binning was not done for TIC 260342097 since only data from two full orbital periods were applied.

\begin{figure}[htbp]
\begin{minipage}[t]{0.49\textwidth}
\centering
\includegraphics[width=9cm,height=6cm]{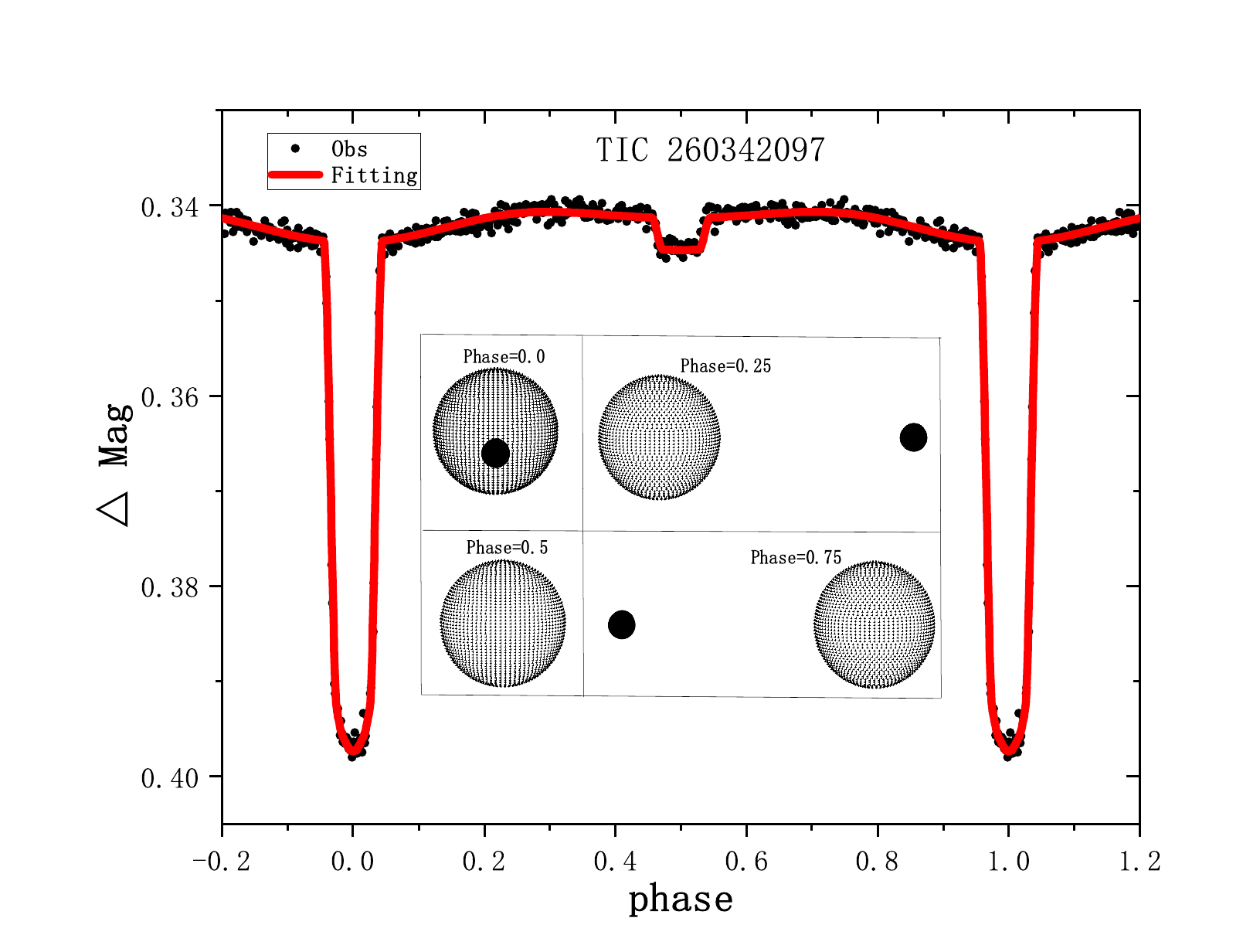}
\end{minipage}
\begin{minipage}[t]{0.49\textwidth}
\centering
\includegraphics[width=9cm,height=6cm]{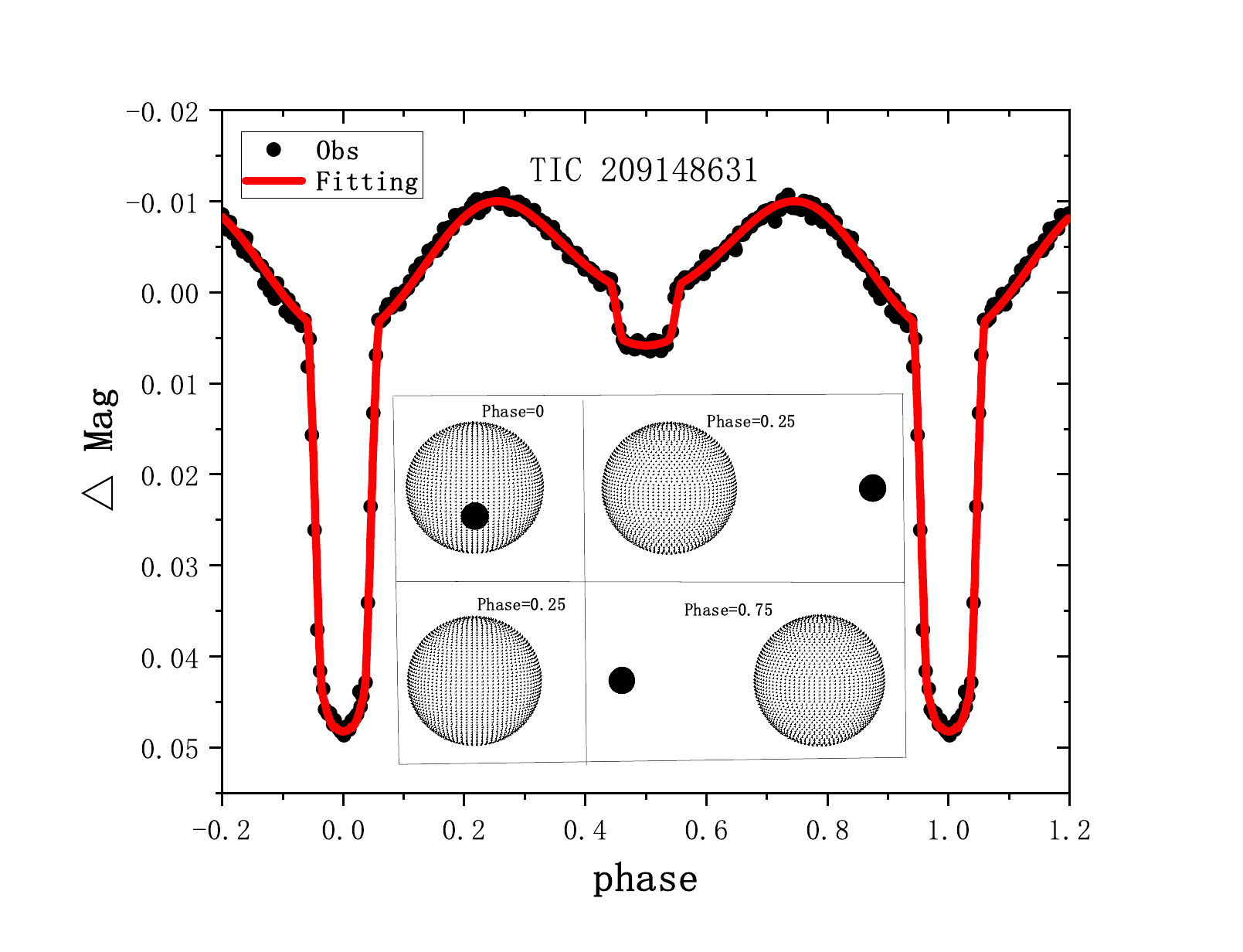}
\end{minipage}
\caption{The observational and fitting light curves of TIC 260342097 and TIC 209148631. In the middle of the panels are their corresponding geometrical structures.}
\label{2fit}
\end{figure}

Before implementing the W-D code to fit the light curves, one should determine the effective temperature of the primary component. For TIC 260342097 and TIC 209148631, we adopted the effective temperature as $10070$ K and $10195$ K from Gaia mission, respectively. Generally, the gravity darkening exponents, $g_{1}$ and $g_{2}$, for the W-D program are taken as 0.32 for the late-type stars with a convective atmosphere, and 1.0 for the hotter stars with a radiative atmosphere in a binary system \citep{1967ZA.....65...89L}. And the bolometric albedos ($A_{1}$, $A_{2}$) are adopted as 0.5 and 1.0 for convective and radiative atmospheres, respectively \citep{1969AcA....19..245R}. The W-D version of 2013 that used in this study enables the automatic calculation of limb-darkening coefficients. The bolometric limb darkening parameters were taken from the table of \cite{1993AJ....106.2096V} and bandpass limb-darkening coefficient specifically for the TESS bandpass were employed (the coefficient files were generously supplied by Professor Van Hamme, who is one of the authors of the W-D code), and  a logarithmic law was utilized. With setting $T_{1}=10070$ K and $T_{1}=10195$ K for TIC 260342097 and TIC 209148631, the fittings start with mode 2 (detached mode) for both light curves. The adjustable parameters to the fits are: the orbital inclination, $i$, the mean temperature of the secondary component, $T_{2}$, the monochromatic luminosity of the primary, $L_{1}$, and the dimensionless potentials of components 1, $\Omega_{1}$ and 2, $\Omega_{2}$. Here we note that the smear effect was taken into account for light curves of TIC 209148631 because of the 30-min exposure. In addition, the light curves, displayed in Figure \ref{2fit}, show that the phase of the shallower eclipses is compatible with 0.5 for both TIC 260342097 and TIC 209148631, suggesting a null eccentricity. Thus, a circular orbit for them was assumed.

\begin{figure}[htbp]
\begin{minipage}[t]{0.49\textwidth}
\centering
\includegraphics[width=10cm]{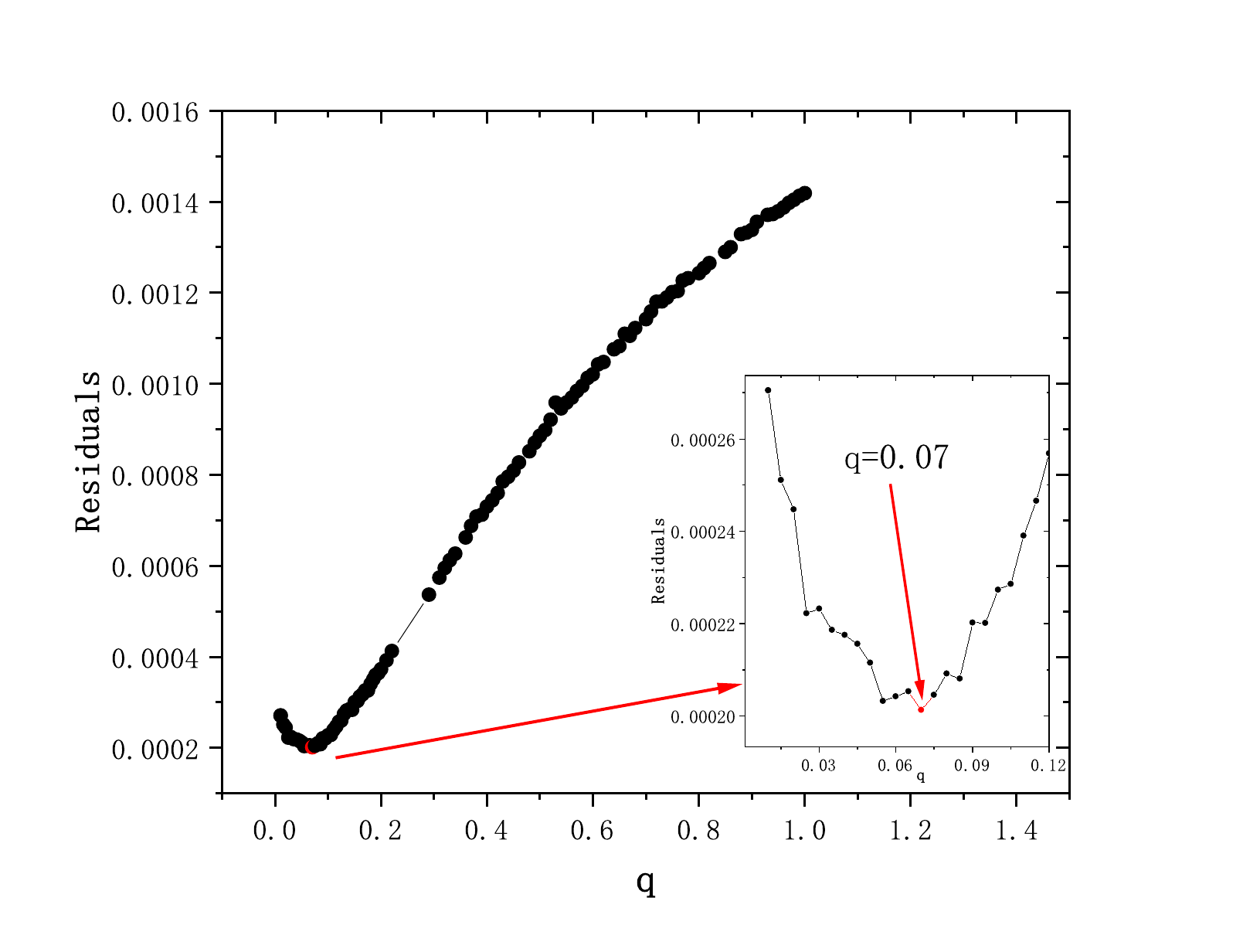}
\end{minipage}
\begin{minipage}[t]{0.49\textwidth}
\centering
\includegraphics[width=10cm]{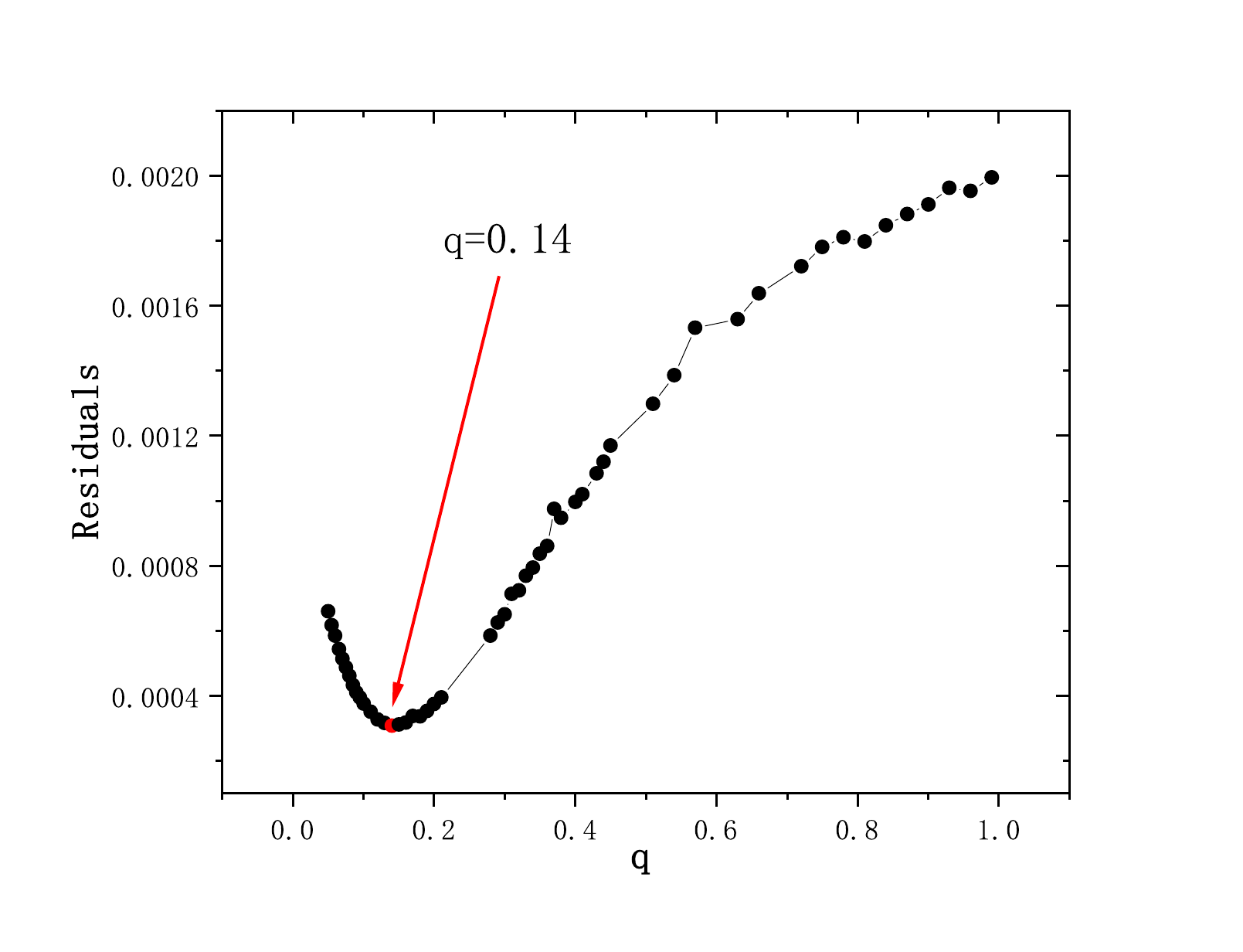}
\end{minipage}
\caption{Residuals - q curves of TIC 260342097 (left) and TIC 209148631 (right).}
\label{q-s}
\end{figure}

In order to find a reliable initial mass ratio, the q-search method was applied, where the solutions based on a group of assumed mass ratio values from 0.01 to 1 were tried with a step of 0.01. The q-search results were plotted in Figure \ref{q-s}, in which the minimum values of residuals were achieved at $q = 0.07$ for TIC 260342097 and $q = 0.14$ for TIC 209148631. Then, we made q an adjustable parameter, setting $q = 0.07$ and $q = 0.14$ as the initial value for TIC 260342097 and TIC 209148631, and performed differential corrections. $L_{3}$ was also set as an adjustable parameter for both light curves, but the values kept zero. The final convergent photometric results were listed in Table \ref{Tab:ele}, which manifest that both systems are detached binary systems. Figure \ref{2fit} presents the observational and fitting light curves of TIC 260342097 and TIC 209148631, along with their corresponding geometrical structures.

\begin{table}
\tiny
\caption{Photometric solutions of TIC 260342097 and TIC 209148631.}\label{Tab:ele}
\begin{center}
\begin{tabular}{cc|cc}
\hline 
 \multicolumn{2}{c} {TIC 260342097}&
\multicolumn{2}{c} {TIC 209148631} \\
\hline
Parameters & Values & Parameters & Values  \\
\hline
$g_{1}$  & 1.0   & $g_{1}$  & 1.0  \\
$g_{2}$  & 0.32  & $g_{2}$  & 0.32 \\
$A_{1}$ & 1.0   & $A_{1}$   & 1.0 \\
$A_{2}$ & 0.5    &$A_{2}$  & 0.5 \\
$i$ (deg)   &  85.226(42)    &$i$ (deg)   &   81.97(18)  \\
$T_{1}$ (K) & 10070         &$T_{1}$ (K)  &   10195   \\
$T_{2}$ (K) & 3468(80)    & $T_{2}$ (K) & 4095(84)  \\
$q(M_{2}/M_{1})$    &  0.0667(19)  & $q(M_{2}/M_{1})$  &  0.1400(33)  \\
$\Omega_{1}$ & 4.3040(87)  & $\Omega_{1}$   &  3.256(19)  \\
$\Omega_{2}$ & 2.805(37)      &  $\Omega_{2}$     &  3.683(63)  \\
$L_{1}/L_{1}+L_{2}$ & 0.99914754(10) & $L_{1}/L_{1}+L_{2}$ & 0.99814800(41)   \\
$L_{2}/L_{1}+L_{2}$ & 0.00085246(10) & $L_{2}/L_{1}+L_{2}$ &0.00185200(41)  \\
$R_{2}/R_{1}$    & 0.2103(54)  & $R_{2}/R_{1}$  &0.1912(39)  \\
$r_{1pole}$ & 0.23596(46)  & $r_{1pole}$ &  0.3203(16) \\
$r_{1point}$ & 0.23803(48)  & $r_{1point}$ &  0.3300(18)  \\
$r_{1side}$ & 0.23765(47) & $r_{1side}$  &  0.3266(18)  \\
$r_{1back}$ & 0.23792(48) & $r_{1back}$ &    0.3287(18)\\
$r_{2pole}$ & 0.0498(24)  & $r_{2pole}$ &  0.0620(24)\\
$r_{2point}$ & 0.0500(25)  & $r_{2point}$ & 0.0622(24)   \\
$r_{2side}$ & 0.0499(24) & $r_{2side}$  &   0.0620(24) \\
$r_{2back}$ &  0.0500(25) & $r_{2back}$ &   0.0622(24) \\
$R_{2}/R_{1}$ &0.2103(54)  &$R_{2}/R_{1}$ &   0.1912(39)\\
Volume filling degree & 0.05901(63)  & Volume filling degree  &  0.2058(28) \\
of star 1 & & of star 1 &\\
Volume filling degree & 0.0201(16)  & Volume filling degree  & 0.0203(13) \\
of star 2 & & of star 2 &\\
$\Sigma$ &0.00018&$\Sigma$&    0.00028\\
\hline
\end{tabular}
\end{center}
\end{table}

\section{Discussion and conclusion} \label{sect:discu}
This study reported the first detail analyses on TIC 260342097 and TIC 209148631. TIC 260342097 was identified as a star and TIC 209148631 was classified as a pulsating variable star in previous studies. However, both targets were discovered as an eclipsing binary system by us when searching for and investigating slowly pulsating B-type stars and $\beta$ Cephei variable stars among O- and B-type massive stars \citep{2023ApJS..268...16S,2023ApJS..265...33S}. We found no pulsation signals about the two targets after thorough analyses. By using the TESS data, we have determined precise orbital periods for both systems: P=1.6161323(2) days for TIC 260342097, and P=2.37857(40) days for TIC 209148631. 

The TESS light curves of both eclipsing binaries were analyzed by using the W-D code \citep{1971ApJ...166..605W,1979ApJ...234.1054W}. It is discovered that both of them are detached binaries with extremely low mass ratios as q=0.0667(19) for TIC 260342097 and q=0.1400(33) for TIC 209148631. Both systems exhibit total eclipses characterizing by a broad and flat minimum, which can be seen clearly in the panels of Figure \ref{2fit}, indicating that the derived photometric solutions for the two eclipsing binaries are reliable \citep{2005Ap&SS.296..221T}. Moreover, by reviewing the Simbad database, there are four stars within $30^{\prime\prime}$ around TIC 260342097 that are at least 7 magnitudes fainter than the target. As for TIC 209148631, two targets are seen near it, which are about 6 magnitudes fainter than the target. These nearby stars have a negligible impact on the flux measured by TESS, which indicates that the light curves analyzed in this work are reliable.

By considering that the primary components for TIC 260342097 and TIC 209148631 are main-sequence (MS) stars, their masses, $M_{1}$, could separately be estimated as $2.40(15) M_{\odot}$ and $2.45(15) M_{\odot}$ based on the effective temperatures $T_{1}$ of $10070$K and $10195$K given by Gaia \citep{2000PhT....53j..77C}. When estimating the masses, the error of $T_{1}$ was adopted as $\pm{324}K$, which is the average uncertainty according to \citet{2018A&A...616A...8A}. With the derived mass ratios, the masses of the secondary components ($M_{2}$) can be calculated as $0.16(1) M_{\odot}$ and $0.34(2) M_{\odot}$. Subsequently, by using the Kepler's Third Law, we computed the semi-major axes (A) for the two binaries as $7.93(16) R_{\odot}$ and $10.56(21) R_{\odot}$, respectively. By combining the relative radii determined from the W-D with the semi-major axes, the mean radii for both components in the two eclipsing binaries were derived as $R_{1}=1.88(4) R_{\odot}$, $R_{2}=0.39(1) R_{\odot}$ for TIC 260342097, and $R_{1}=3.45(7) R_{\odot}$, $R_{2}=0.65(2) R_{\odot}$ for TIC 209148631. Finally, their luminosities were estimated as $L_{1}=32.26\pm{5.30} L_{\odot}$, $L_{2}=0.020(4) L_{\odot}$ for TIC 260342097, and $L_{1}=113.55\pm{18.42} L_{\odot}$, $L_{2}=0.11(2) L_{\odot}$ for TIC 209148631, respectively. Here we note that all uncertainties in the parentheses are on basis of error transfer formulas. However, The errors listed in Table \ref{Tab:ele} are the standard deviations directly calculated by the W-D code, which, for the nonlinear situations, are not ideally correct. The real uncertainties may be three or five times larger (e.g., \citealt{2015AJ....149..148L}). In addition, the two eclipsing binaries are bright and short-period, making them good candidates for radial velocity (RV) observations. Based on the estimated parameters, the semi-amplitude of the primary components, $K_{1}$, could be estimated as 15.5(5) km/s and 27.3(8) km/s for TIC 260342097 and TIC 209148631, respectively. The estimated semi-amplitude may serve as a reference for planning RV follow-up.

\begin{figure}[htbp]
\begin{minipage}[t]{0.45\textwidth}
\centering
\includegraphics[width=8.5cm]{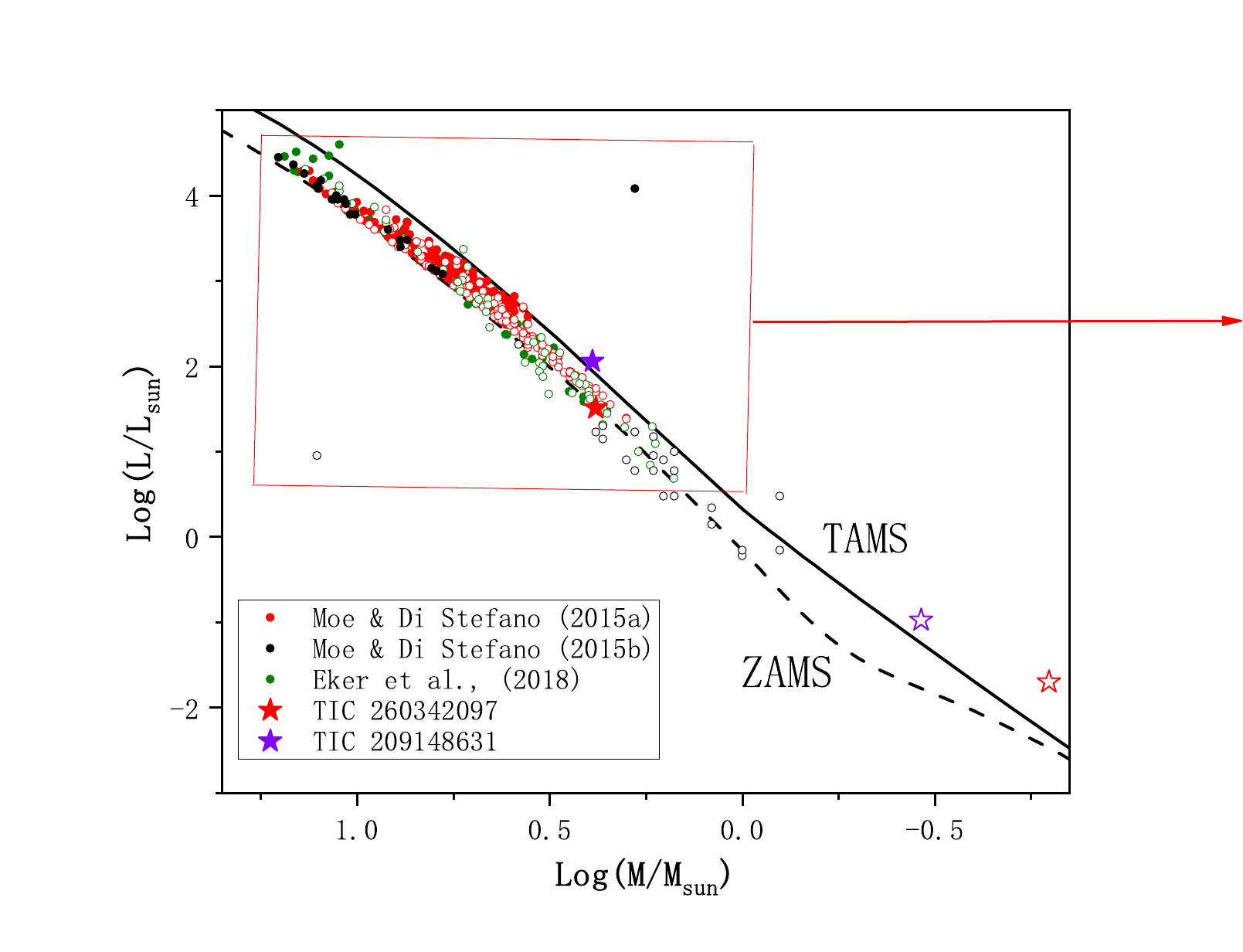}
\end{minipage}
\begin{minipage}[t]{0.45\textwidth}
\centering
\includegraphics[width=8.6cm]{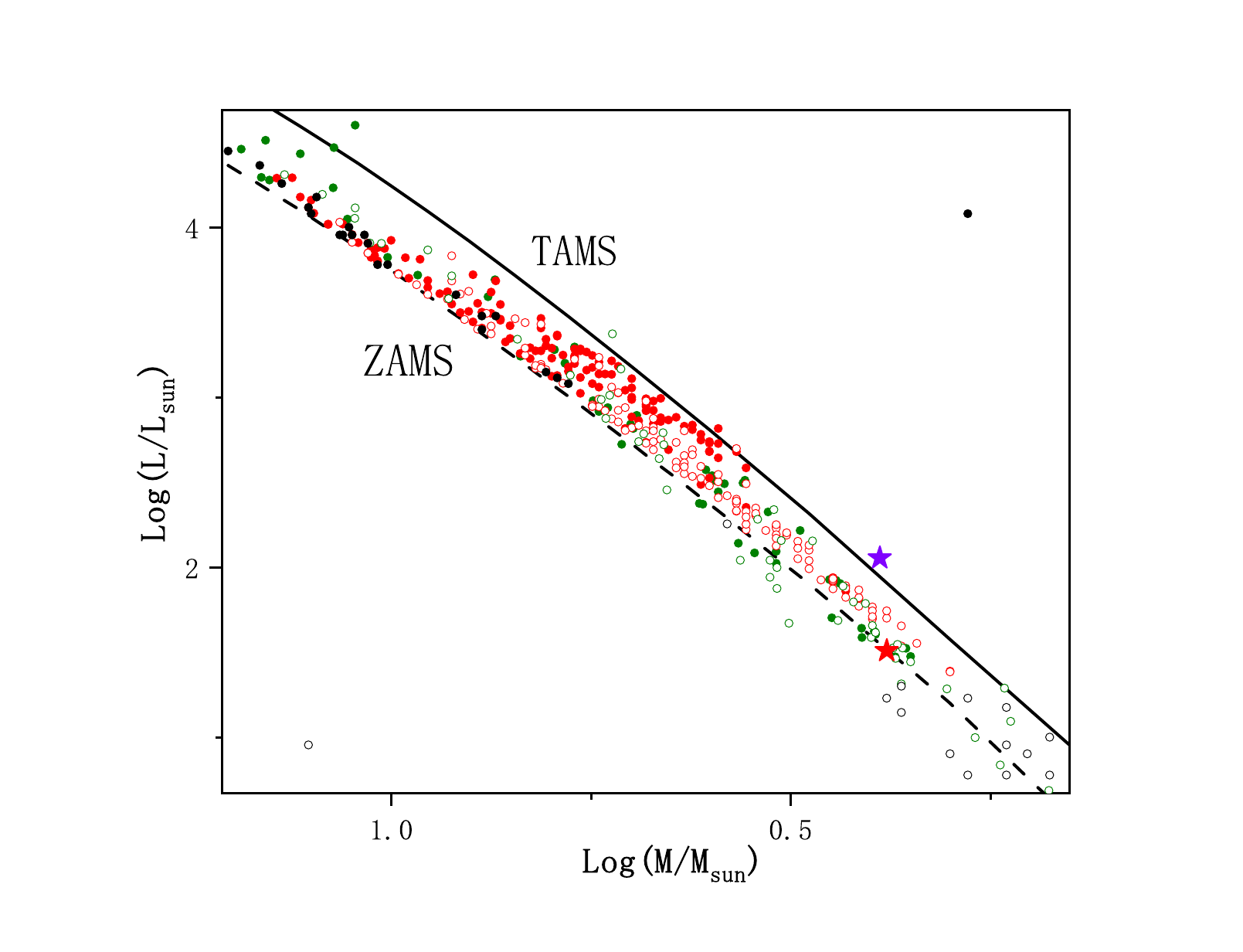}
\end{minipage}

\begin{minipage}[t]{0.45\textwidth}
\centering
\includegraphics[width=8.5cm]{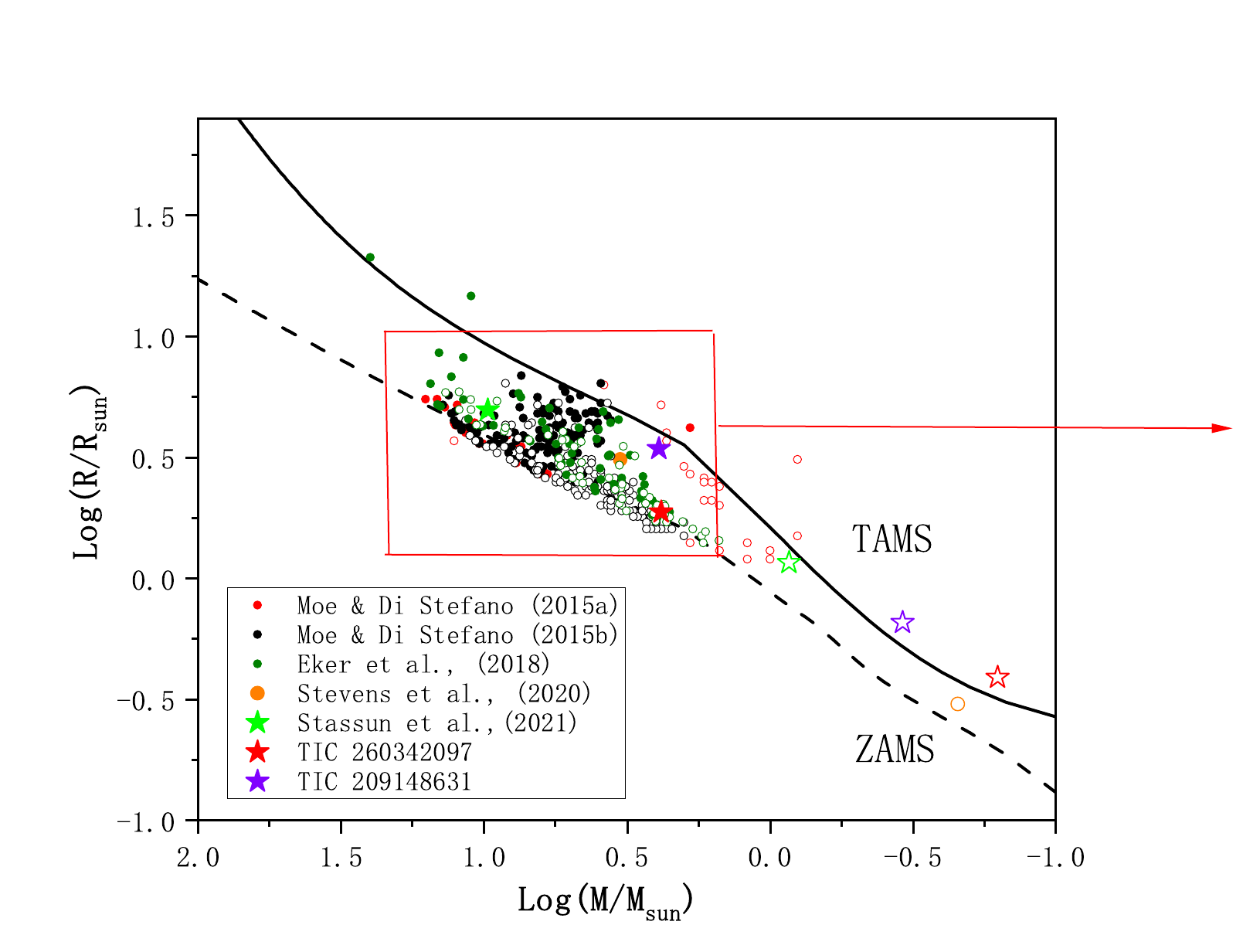}
\end{minipage}
\begin{minipage}[t]{0.45\textwidth}
\centering
\includegraphics[width=8.5cm]{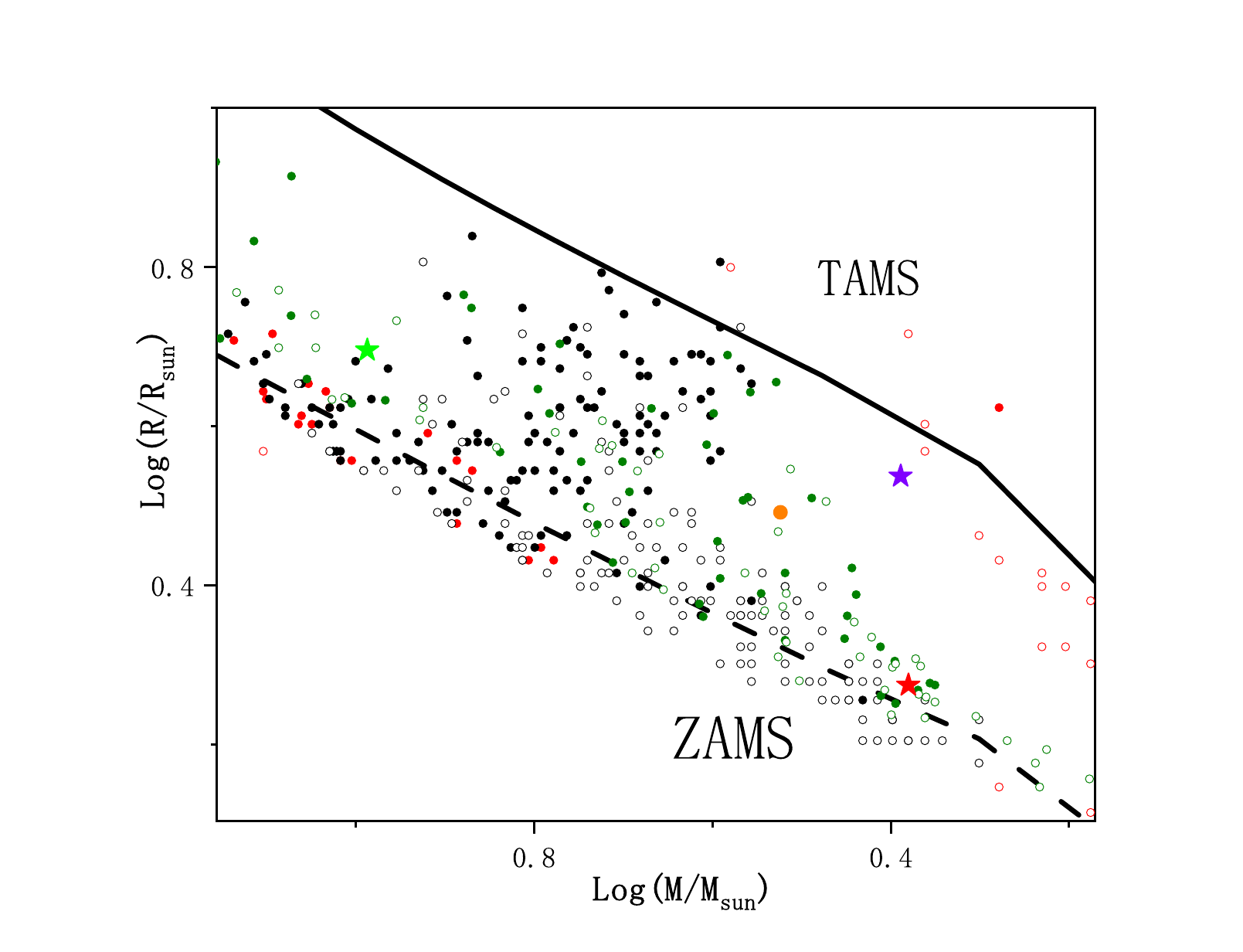}
\end{minipage}

\makebox[\textwidth][c]{%
\begin{minipage}[t]{0.7\textwidth}
\centering
\includegraphics[width=10.5cm]{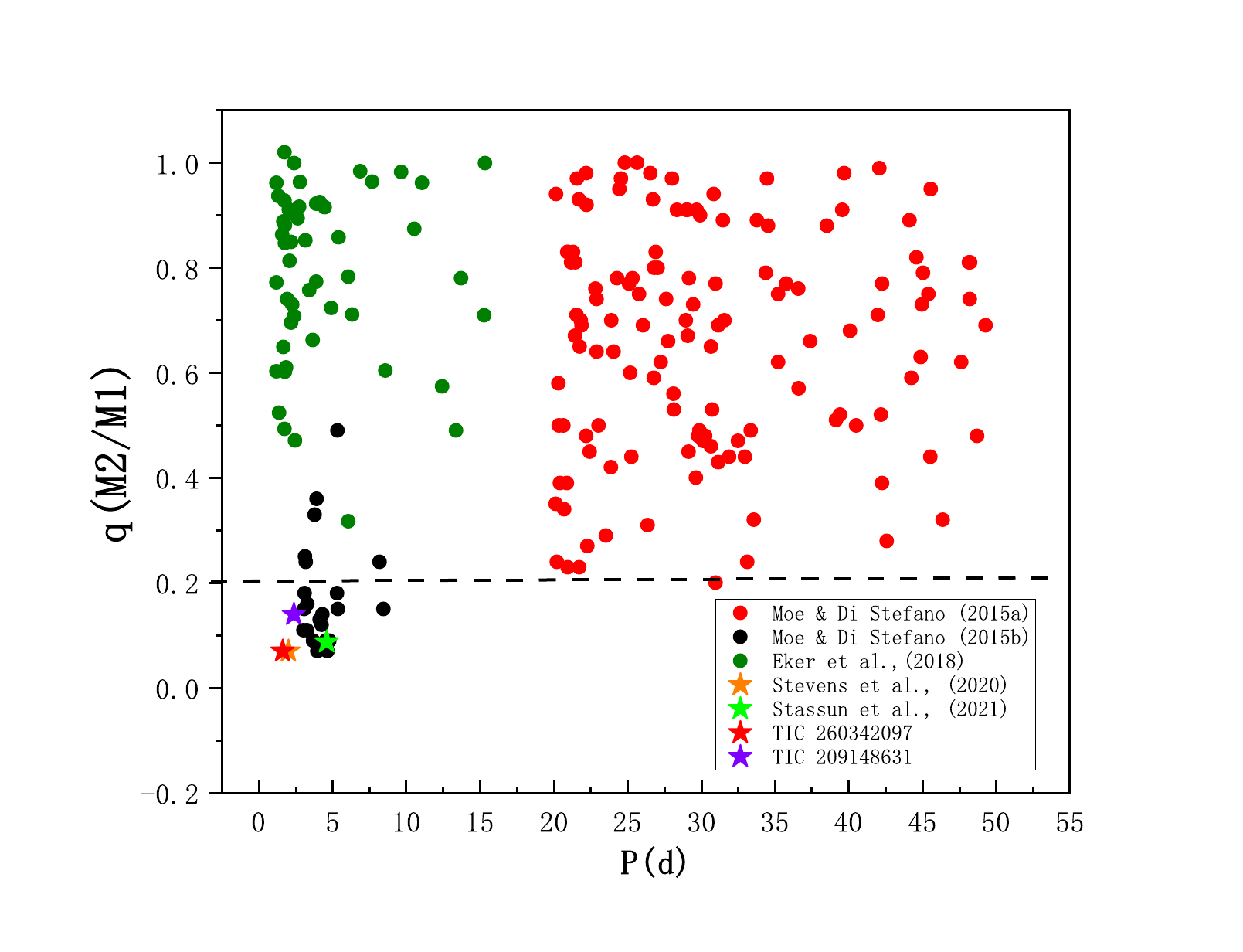}
\end{minipage}}
\caption{Positions of TIC 260342097 and TIC 209148631 on the mass-luminosity (M-L, upper panel), mass-radius (M-R, middle panel), and P-q distributions (bottom panle). The top right panel shows the enlarged section of the M-L diagram circled in a red box, while enlarged portion of the M-R diagram is placed on the middle right panle. Solid symbols present the primaries, while the same symbols of hollow denote the low-mass companions. Also displayed in the panels are the samples from \citet{2015ApJ...810...61M,2015ApJ...801..113M}, \citet{2018MNRAS.479.5491E},
 \citet{2020MNRAS.499.3775S} and \citet{2021ApJ...910..133S}. The dashed line in the bottom panel refers to systems with mass ratio of 0.2.}
\label{HR}
\end{figure}

\begin{figure}[htbp]
\centering
\begin{minipage}[t]{0.45\textwidth}
\centering
\includegraphics[width=8.5cm]{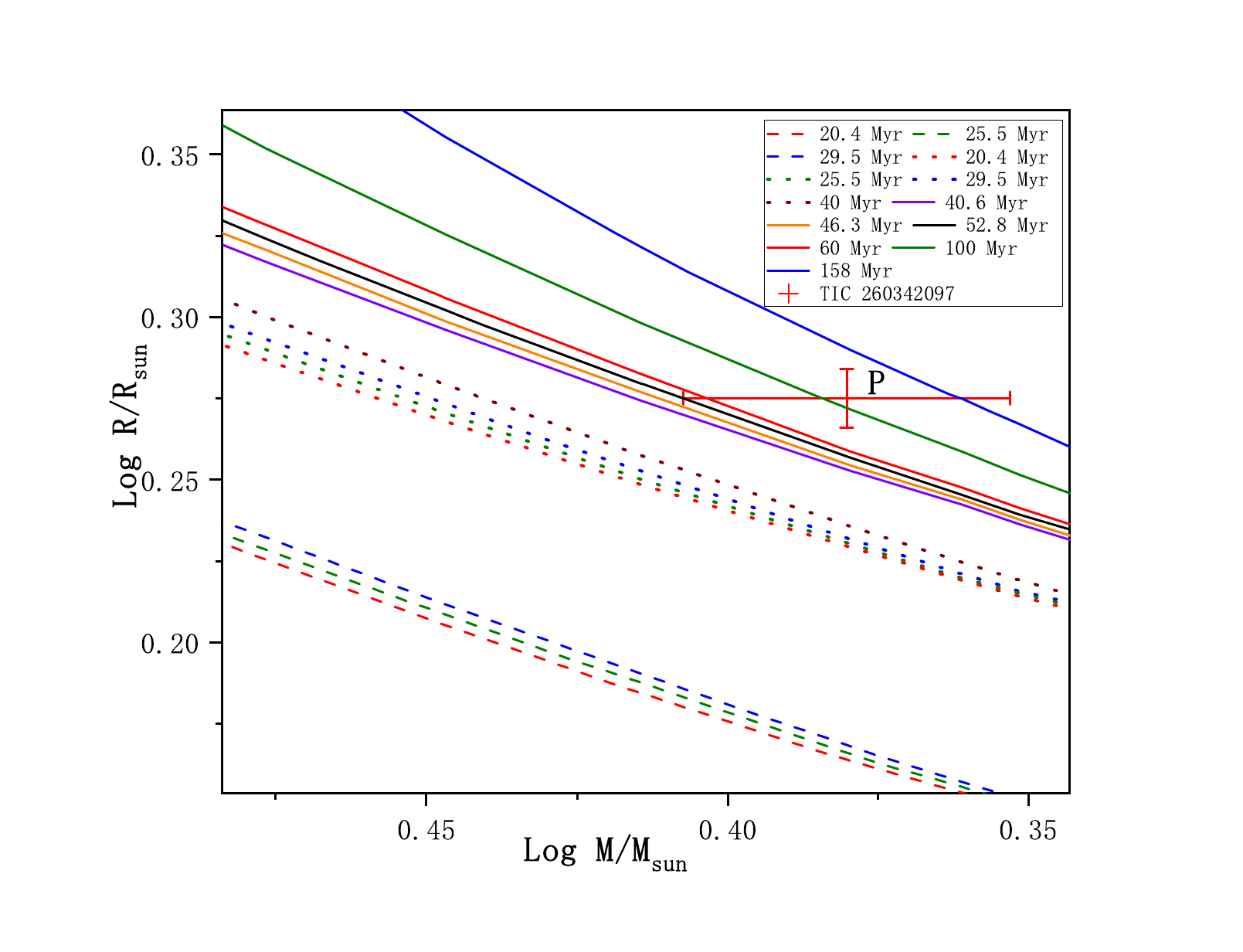}
\end{minipage}
\begin{minipage}[t]{0.45\textwidth}
\centering
\includegraphics[width=8.5cm]{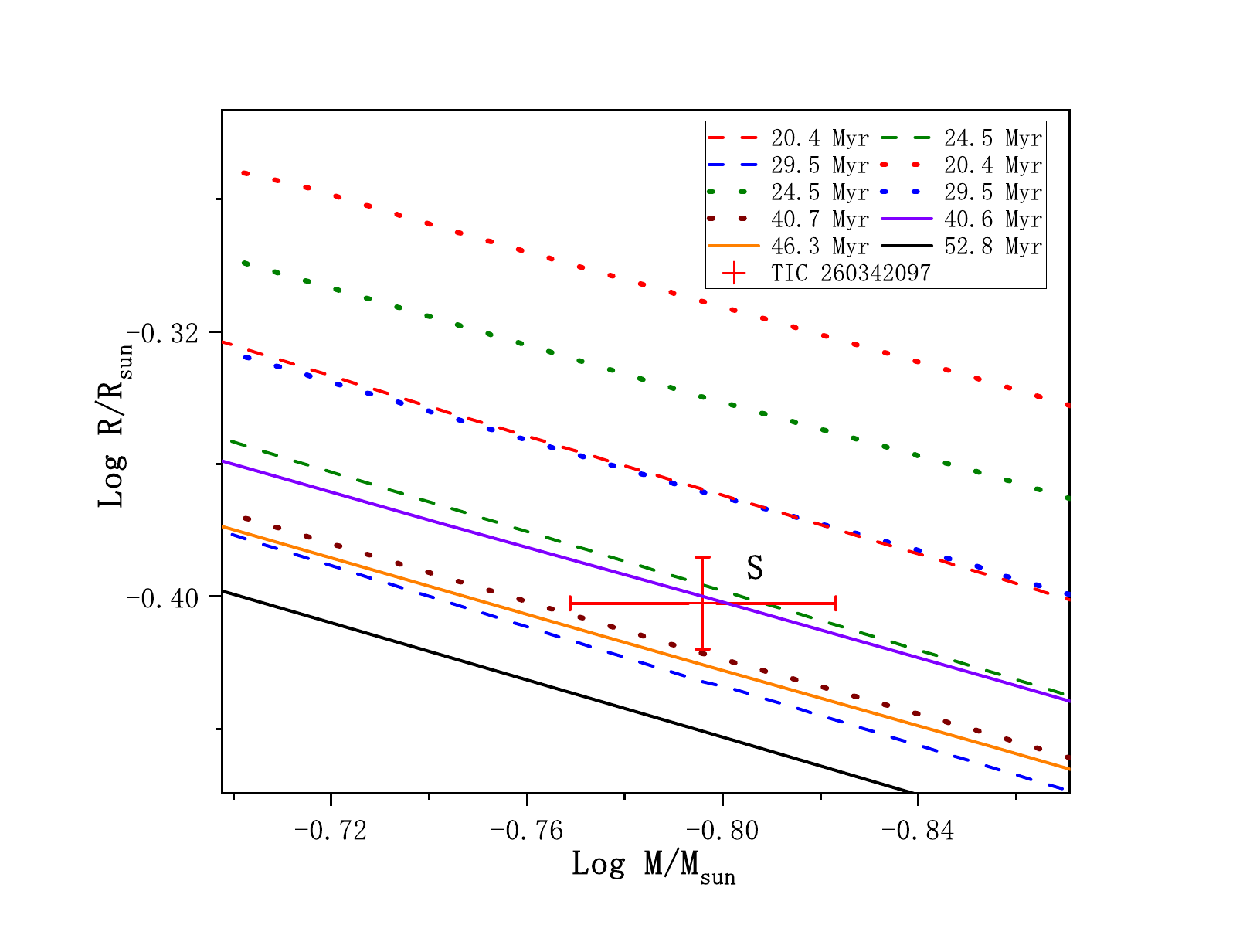}
\end{minipage}
\caption{Mass–radius diagrams for the primary (P, left panel) and secondary (S, right panel) components (with error bars) of TIC 260342097 compared to PARSEC isochrones at [M/H] = -0.5 (dashed lines), solar metallicity (dotted), and [M/H] =0.134 (solid).}
\label{isochrone}
\end{figure}

Recently, \citet{2018MNRAS.479.5491E} compiled parameters of 293 detached binary stars, including 52 B-type binary systems. The mass-luminosity, mass-radius, and period-mass ratio relations of these 52 B-type systems are displayed in Figure \ref{HR}, where the luminosities were computed by us with the effective temperatures and the radii provided by \citet{2018MNRAS.479.5491E}. Also shown in the panels are 152 B-type binary stars in Large Magellanic Cloud (LMC) that were investigated by \citet{2015ApJ...810...61M,2015ApJ...801..113M}, and two samples from \citet{2020MNRAS.499.3775S} and \citet{2021ApJ...910..133S}. The companions of the B-type primaries in those binaries are main-sequence (MS) or pre-MS stars. For comparison, the two binaries, TIC 260342097 and TIC 209148631, are also plotted in Figure \ref{HR}. As with the binary system investigated by \citet{2020MNRAS.499.3775S}, TIC 260342097 has an extremely low mass ratio as 0.07. Both of them have the shortest orbital periods and lowest mass ratio. As displayed in the upper and middle panels of Figure \ref{HR}, the primary of TIC 260342097 is close to the Zero Age Main Sequence (ZAMS), and that of TIC 209148631 is approaching the Terminal Age Main Sequence (TAMS), while the secondary components exceed the TAMS, indicating that the secondaries are “over-luminous” and “over-size” for their companions. The explanation is that they are lying in Pre-MS stage, like the samples from \citet{2015ApJ...801..113M}, \citet{2020MNRAS.499.3775S} and \citet{2021ApJ...910..133S}. The two binary systems may be two new nascent eclipsing binaries with extremely low mass ratios. 
Especially, TIC 260342097 is a member of the open star cluster BH\_164 \citep{2018A&A...618A..93C}. \citet{2020AJ....160..279K} provided the age of cluster BH\_164 as log(age) = 7.39(8) yr, or an age of 20-30 Myr (million years). While \citet{2021MNRAS.504..356D} reported the age of cluster BH\_164 to be log(age) = 7.666(57) yr, corresponding to an age span of roughly 40 to 52 Myr, and determined its metallicity as $[Fe/H]=0.134(52)$. An analysis using the PAdova and TRieste Stellar Evolution Code (PARSEC) isochrones \citep{2012MNRAS.427..127B}, as depicted in Figure \ref{isochrone}, compares the components of TIC 260342097 on a mass-radius diagram, considering metallicities of [M/H]=-0.5 (dashed lines), solar (dotted), and [M/H] =0.134 (solid), where the last is an approximation of the metallicity from \citet{2021MNRAS.504..356D}. Figure \ref{isochrone} (right panel) clearly illustrates that the secondary component of TIC 260342097 is presently situated in the Pre-MS stage, contracting towards the MS, whereas the primary component (left panel) is not lying in the Pre-MS stage. And when [M/H] = -0.5, the secondary component is consistent with the age of cluster BH\_164 estimated by \citet{2020AJ....160..279K} (about 25 Myr). However, when it is at solar metallicity or [M/H] =0.134, the age of the secondary component is about 40 Myr, which is close to the cluster's age given by \citet{2021MNRAS.504..356D}.

It is well known that the binary system fraction of main sequence increases with the mass of the primary, consequently, O- and B-type massive stars predominantly exist in binaries (including multiple systems). However, massive binaries with low mass ratios are less explored owing to their detection remains a challenge given that the luminosity contrast between the two components, where the low-mass components in such systems contribute such small flux to total flux that it can hardly be detected with ground-based telescopes. As shown in the bottom panel of Figure \ref{HR}, among the 208 B-type binary stars, only a few percent of systems have mass ratio below 0.2. They may be formed from fragmentation of the circumstellar disk \citep{2006MNRAS.373.1563K,2013AJ....145....3G}. \citet{2011ApJ...730...32S} pointed out that the episodic accretion leads to the formation of low-mass stars, brown dwarfs, and planetary-mass objects by disk fragmentation. The detection of two B-type binaries with extremely low mass ratios are very important for understating the formation and evolution of massive binaries with low mass companions. They may be the progenitors of cataclysmic variables \citep{2023MNRAS.525.1641N}. In the future, TESS will detect more such binary systems.

We are very grateful to the referee for his/her helpful comments and suggestions, which have greatly contributed to improving our manuscript. This work is supported by National Key R\&D Program of China (grant No.2022YFE0116800) and National Natural Science Foundation of China (grant Nos.11933008 and 11922306). The continuous photometric data used in this study are collected by the TESS mission. Funding for the TESS mission is provided by NASA Science Mission Directorate. We really appreciate the TESS team for supporting of this work. This work also makes use of results from the European Space Agency (ESA) space mission Gaia. Gaia data are being processed by the Gaia Data Processing and Analysis Consortium (DPAC). Funding for the DPAC is provided by national institutions, in particular the institutions participating in the Gaia MultiLateral Agreement (MLA).

\bibliography{sample631}{}
\bibliographystyle{aasjournal}



\end{document}